\documentclass[a4paper, Garamond, 12pt]{article}

\usepackage[latin1]{inputenc}
\usepackage{graphicx, amssymb, amsfonts, amsmath, mathrsfs}
\usepackage[english]{babel}
\usepackage[official]{eurosym}
\usepackage{caption}[2008/04/01]
\usepackage{newcent}
\usepackage{rotating}
\usepackage{setspace}
\usepackage{tabularx}
\usepackage[flushmargin]{footmisc} 
\usepackage{color}
\usepackage{lscape,array,hhline,pstricks,fancyhdr}
\usepackage{lineno}
\usepackage[hscale=1,vscale=1]{geometry}
\usepackage{geometry}
\usepackage{natbib}
\usepackage{url}
\usepackage{threeparttable}

\title{Are Chinese transport policies effective? A new perspective from direct pollution rebound effect, and empirical evidence from road transport sector}

\author{Lu-Yi QIU $^{1}$  and Ling-Yun HE $^{1, 2, 3,}$\footnote{Dr. HE is the corresponding author. Dr. HE is a full professor of energy economics and environmental policies. QIU is a Ph.D. candidate supervised by Dr. HE.  The authors contribute equally in the project. HE conceived the whole project. QIU calculated and analysed the results under Dr. HE's supervision. HE and QIU co-wrote the manuscript. The authors would like to thank Dr. YANG Sheng, Dr. CHEN Su-Mei, XU Feng, LIU Li, OU Jia-Jia, WEI Wei, and all other colleagues from both China Agricultural University and JiNan University, for all their warm helps, constructive suggestions and pertinent comments. This project is supported by the National Natural Science Foundation of China (Grant Nos. 71273261 and 71573258), and China National Social Science Foundation (No. 15ZDA054).}       \\ \small 1. College of Economics and Management, China Agricultural University,\\ \small Beijing 100083, China\\\small 2. Institute of Resource, Environment and Sustainable Development Research, \\ \small JiNan University, Guangzhou 510632, China \\ \small 3. School of Economics, JiNan University, Guangzhou 510632, China\\ \small 4. School of Economics and Management, Nanjing University of Information Science and Technology,\\ \small Nanjing 210044, China\\
 \small * Corresponding author.
\\ \small Emails: louieq@126.com; lyhe@amss.ac.cn}

\date{\small Submitted on \today}

\begin{document}

\maketitle

\newpage
\begin{abstract}
The air pollution has become a serious challenge in China. Emissions from motor vehicles have been found as one main source of air pollution. Although the Chinese government has taken numerous policies to mitigate the harmful emissions from road transport sector, it is still uncertain for both policy makers and researchers to know to what extent the policies are effective in the short and long terms. Inspired by the concept and empirical results from current literature on energy rebound effect (ERE), we first propose a new concept of  ``pollution rebound effect" (PRE). Then, we estimate  direct air PRE as a measure for the effectiveness of the policies of reducing air pollution from transport sector based on time-series data from the period 1986 -- 2014. We find that the short-term direct air PRE is $-1.4105$, and the corresponding long-run PRE is $-1.246$. The negative results indicate that the direct air PRE does not exist in road passenger transport sector in China, either in the short term or in the long term during the period 1986--2014. This implies that the Chinese transport policies are effective in terms of harmful emissions reduction in the transport sector. This research, to the best of our knowledge, is the first attempt to quantify the effectiveness of the transport policies in the transitional China.
\end{abstract}

\newpage
\begin{onehalfspace}

\section{Introduction}\label{section_intro}

\begin{figure}[h!]
\centering
\includegraphics[width=14cm, height=7.5cm]{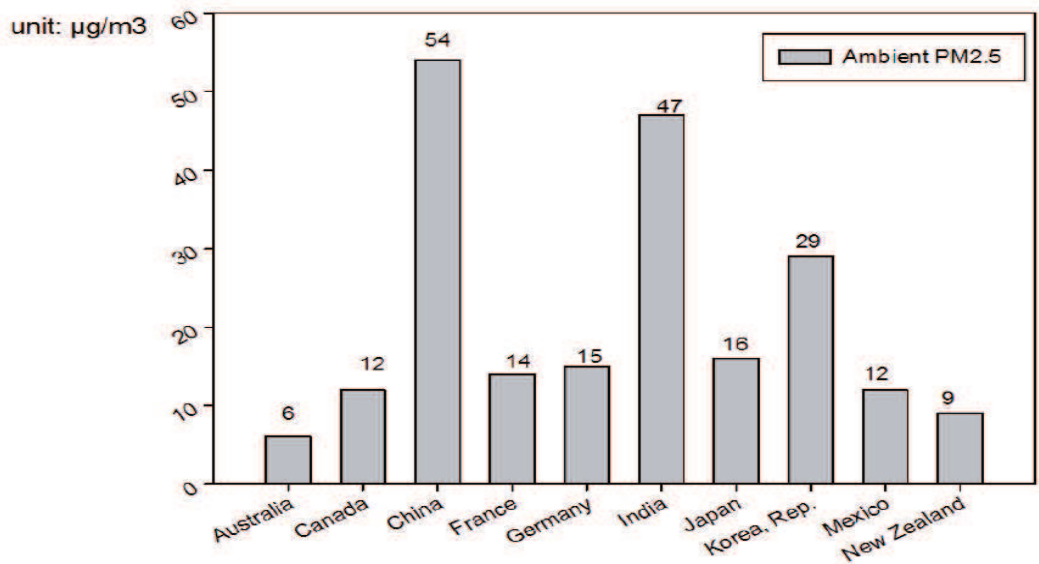}
\caption{ Concentration of ambient PM2.5 in different countries, 2013 \\ (Data source: World Development Indicators 2016)}
\label{fig1}
\end{figure}

China is facing a serious environment problem, especially air pollution resulting from the rapid economic growth. According to one study of World Bank (2007), twelve of the twenty most polluted cities in the world are located in China. This ranking is based on ambient concentrations of particulate matter less than 10 $\mu m$ in diameter. The ambient concentration of $\mathrm{PM2.5}$ in China is the most polluted in the world based on the report of World Bank (2016) (see Fig. \ref{fig1}). The State of Environment (SOE) Report of 2016 indicates\footnote{Ministry of Environmental Protection of the People's Republic of China, 2016. \url{http://www.mep.gov.cn/gkml/hbb/qt/201604/t20160421_335390.htm}}: ``Among the 338 prefecture-level cities, there are eighty percent whose air quality exceed the standard, and 45 cities exceed the annual average concentration of fine particulate matter more than doubled in 2015." Serious air pollution has severe effects on human health, increasing the risk of lung cancer, respiratory and cardiovascular diseases (Kunzli et al., 2000; Hoek et al., 2002; Samet, 2007; Beelen et al., 2008; Brunekreef et al., 2009; Weichenthal et al., 2011), which also increases the residents' medical cost (Yang et al., 2013; Chen and He, 2014; Yang and He, 2016). Now more and more public pay attention to air quality, which poses more pressure on Chinese government to make scientific and feasible policies to balance between economic development and environment problems.

\begin{figure}[h!]
\centering
\includegraphics[width=14cm, height=8cm]{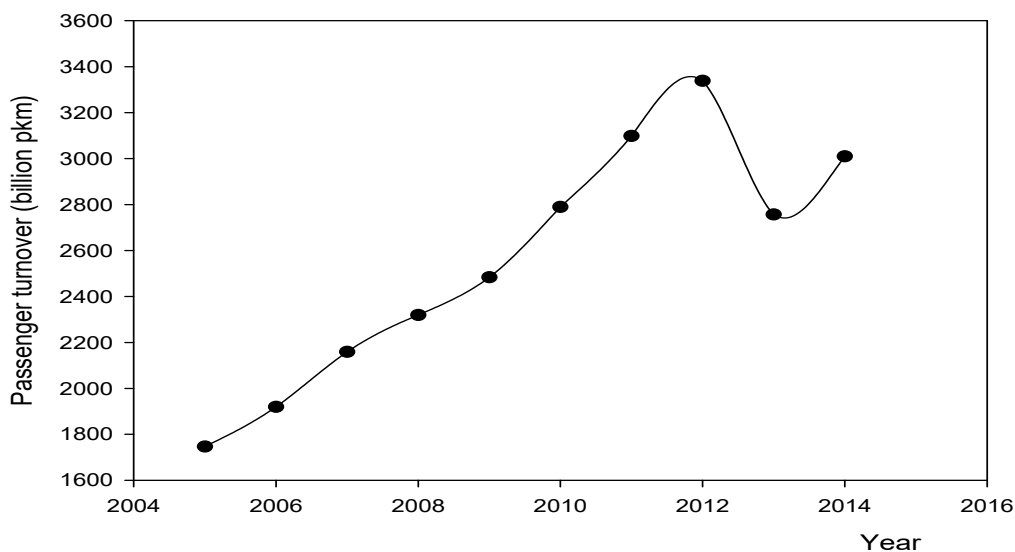}
\caption{Passenger turnover in China, 2005-2014.\\  (Data source: National Bureau of Statistics of the People's Republic of China)}
\label{fig2}
\end{figure}

To control and decrease air pollution, it is necessary to figure out the main sources of air pollution. According to ``China Vehicle Environmental Management Annual Report, 2016", motor vehicles have been one main source of air pollution in China\footnote{Ministry of Environmental Protection of the People's Republic of China, 2016. \url{http://www.mep.gov.cn/gkml/hbb/qt/201606/t20160602_353152.htm}}. The transport sector is a major area that policymakers should pay more attention. The last decade has witnessed a dramatic increase of the vehicles stock, causing the rapidly growing travel demand of the Chinese residents. China's passenger turnover has risen from 1746.67 billion passenger-kilometers (pkm) in 2005 to 3009.74 billion pkm in 2014 (National Bureau of Statistics of China, 2006--2015)(see Fig. \ref{fig2}), resulting in serious polluted air emissions. The total vehicle emissions in China reached to 45.32 million tons in 2015. Specifically, the emissions of $\mathrm{CO}$, $\mathrm{HC}$, $\mathrm{NO_{x}}$ and $\mathrm{PM}$ from vehicles were 34.61, 4.30, 5.85 and 0.56 million tons, respectively\footnote{ Data source: Ministry of Environmental Protection of the People's Republic of China, 2016. \url{http://www.mep.gov.cn/gkml/hbb/qt/201606/t20160602_353152.htm}}. The transport sector has been a major field of harmful emissions reduction.

Chinese government has implemented several laws and policies to deal with the serious air pollution. The law ``Prevention and Control of Air Pollution" was introduced in 1987 by the National People's Congress and its Standing Committee. A wide range of regulations, decisions, orders and quality standards have been issued. For example, the State Council promulgated ``Atmospheric Pollution Prevention Action Plan" in 2013. In 2004, the mandatory fuel economy standard for passenger vehicles was launched and the first, second and third phases were implemented in 2005, 2008 and 2010, respectively. In some megacities (e.g. Beijing, Shanghai and Guangzhou), regulatory policies imposed on vehicle usage, as well as car ownership. Chinese government also takes various policies to encourage and popularize the research and development of new energy vehicles, like tax preferential policy, technology innovation policy and financial subsidy policy et al..

Due to the efforts of Chinese government, the environment quality has been improved. But China's environment protection is still lagging behind in economic and social development. Environment carrying capacity has been reached or closed to the ceiling\footnote{Ministry of Environmental Protection of the People's Republic of China, 2016. \url{http://www.mep.gov.cn/gkml/hbb/qt/201604/t20160421_335390.htm}}. To accurately understand the actual effect of these policies, a measure of rebound effect (RE) is necessary, which can provide useful information about the effectiveness of the policies for the policymakers.

The rebound effect is initially proposed by Jevons (1866). It is generally acknowledged that when technological progress causes an increase in efficiency by 1\%, a reduction in energy consumption obtaining the same products by 1\% is expected, whereas the actual reduction may be below 1\%. Studies have identified three main types of rebound effects (RE) (e.g., Berkhout et al., 2000; Greening et al., 2000; Frondel et al., 2008; Sorrell and Dimitropoulos, 2008): direct rebound effect, indirect rebound effect and macro-level rebound effect.

Direct rebound effect is limited to a single energy service or a single sector. With the improvement of energy efficiency, energy consumption is not reduced to the expected level in theory because of the decline in the cost of energy product or energy service and the increase in consumers' energy demand. The issue of pollution rebound effect (PRE) in transport sector also relates to the improvement in energy efficiency. The government encourages to improve the energy efficiency of vehicles to reduce harmful emissions and save energy from the travel. However, fuel-efficient vehicles make energy services cheaper, thereby encouraging the increased consumption of those services. For instance, consumers may choose to drive farther and/or more often following the purchase of a fuel-efficient vehicle because the operating cost per kilometer has fallen. It may offset some savings because of fuel efficiency improvement. So there will be rebound effect of the fuel consumption, which results in the harmful emissions also appearing rebound effect. Indirect rebound effect measures the reallocation of energy savings to spending on other goods and services that also require energy. Macro-level  rebound effect refers to the impact of energy efficiency improvement on the entire economy. This paper focuses on the direct air pollution rebound effect from transport sector.

Based on the definition of rebound effect in energy consumption, we firstly define the pollution rebound effect (PRE), namely,

 \begin{equation} \label{eq:r1}
 PRE(\%)=\frac{PlanedEmissionReductions-ActualEmissionReductions}{PlanedEmissionReductions}\times 100\%
 \end{equation}

According to the magnitude, PRE can be classified into five categories, which represent different policy effects (see Table \ref{tab1}). When the size of PRE is greater than 0, it represents that pollution rebound effect exists. 0\textless \text{PRE}\textless 1 means that the polices are partially effective, not fully achieving the goal of pollution emission reductions. PRE=1 means that the policies are completely ineffective and PRE\textgreater 1 is the worst, implying that the polices not only decrease the harmful emissions, but also increase the emissions. When the size of PRE is less than 0 or equal to 0, it represents that pollution rebound effect does not exist. PRE=0 means that the policies are effective, which fully achieve the goal of emission reductions. PRE\textless 0 is the best to the policymakers, which means the actual reductions are more than the planed reductions.

\begin{table}[h!]
\centering
\begin{threeparttable}[b]
\captionsetup{font={scriptsize}}
\caption{Categories of PRE based on the size} \label{tab1}
\scriptsize
    \begin{tabular}{p{2.5cm}p{2.5cm}p{3cm}}
    \hline
     Size                                 &Existence          &Policy implication         \\ \hline
     PRE\textgreater 1                    &Yes                &Negative effect        \\
     PRE$=$1                              &Yes                &Completely ineffective             \\
     0\textless PRE\textless 1            &Yes                &Partially ineffective       \\
     PRE$=$0                              &No                 &Fully effective                \\
     PRE\textless 0                       &No                 &Positive effect        \\  \hline
   \end{tabular}
\end{threeparttable}
\end{table}

Some studies have examined the rebound effect in transport sector, but most are focus on the fuel consumption. Sorrell et al. (2009) provide a review of studies that include transportation and energy in general. They report that for personal automotive transport, in OECD countries, the mean value of the long-run direct rebound effect is likely to be less than 30\% and may be closer to 10\% for transport. Small and Van Dender (2007), examining motor vehicle transportation in the US, estimate the short and long-run rebound effect of 4.5\% and 22.2\%, respectively. Barla et al. (2009) present estimates of the rebound effect for the Canadian light-duty vehicle fleet. Their results imply a rebound effect of 8\% in the short term and a little less than 20\% in the long term. Hymel and Small (2015), using panel data on U.S. states, confirm the earlier finding of a rebound effect that declines in magnitude with income, but they also find an upward shift in its magnitude of about 0.025 during the years 2003--2009. Wang et al. (2012) and Zhang et al. (2015) also estimate the direct rebound effect for passenger transport in China. Although all the current studies conclude that the rebound effect exists in fuel consumption for transport sector, the range of the magnitude is very different. For example, Wang et al. (2012) estimate the direct rebound effect for passenger transport in urban China, finding that the average rebound effect for passenger transport by urban households is around 96\%. Zhang et al. (2015), analyzing road passenger transport in the whole country, eastern, central and western China, reveal that the short-term and long-term direct rebound effects of the whole country are 25.53\% and 26.56\% on average, respectively. Furthermore, we can not find any study on the pollution rebound effect.

In summary, it is necessary to explore that whether direct air pollution rebound effect exist in the road passenger transport in China. To the best of our knowledge, this paper is the first attempt in current literature to evaluate the direct air pollution rebound effect. The results can provide useful information for policy makers to understand the effectiveness of the policies, which aim to reduce harmful emissions of transport sector.

The remaining of this paper is structured as follows. Section 2 introduces the methods used to estimate the rebound effect of air pollution as well as data definitions. Section 3 present the empirical results and detailed discussions. Finally, Section 4 summarizes our results and offers some policy implications.

\section{Methods and data}\label{section_theory}

In this paper we explore the existence of direct air pollution rebound effect for the road passenger transport sector in China during 1986--2014. To estimate the direct air pollution rebound effect, we firstly need to calculate the emissions reduction from transport sector based on the definition of PRE (see Eq. (\ref{eq:r1})). Here, we have an assumption that the emission factor remains unchanged. The details are provided in Section \ref{section_RE}. So we can directly calculate the fuel consumption reduction. According the analysis, we find that we can estimate the fuel rebound effect, which is equal to the PRE. Following the definition by Khazzoom (1980) and Berkhout et al. (2000), we can calculate the rebound effect according to the elasticity of fuel consumption with respect to fuel efficiency. In this paper, we use the elasticity of vehicle kilometers (VKM) with respect to fuel price as a proxy.

\subsection{The direct air pollution rebound effect}\label{section_RE}

Like the estimation in the literature by Chen and He (2014), we calculate the air pollutant emissions of transport by the following equation:

\begin{equation} \label{eq:r2}
 H_i=F_m\times A_{m,i}
\end{equation}

where $H$ refers to the traffic-related harmful gases emissions, $F$ the consumption of the fuel from transport, $A$ the emission factor; $m$, $i$ refer to the fuel type and pollutant type, respectively.

In the last thirty years, the technology of vehicles does not have revolutionary innovation, which still mainly burn fuel. The emissions still have strong relationship with fuel consumption. So in this paper we suppose the emission factor remains unchanged. When the fuel consumption has rebound effect, then the air pollution rebound effect will occur. The both two are equal.

According to the definition of rebound effect (e.g. Khazzoom, 1980; Berkhout et al., 2000), the estimation can be calculated by the following equation:

\begin{equation} \label{eq:r3}
 RE=\eta_E(F)+1
\end{equation}
where $RE$ is the rebound effect; $F$ is the consumption of fuel; $E$ is fuel efficiency; $\eta_E(F)$ refers to the elasticity of fuel consumption with respect to fuel efficiency.

Following the previous study by Odeck and Johansen (2016), the relationship between the elasticity of fuel consumption with respect to fuel efficiency and elasticity of vehicles kilometers traveled (VKM) demand with respect to fuel price is like that:

\begin{equation} \label{eq:r4}
 \eta_E(F)=-\eta_P(VKM)-1
\end{equation}

where $\eta_P(VKM)$ is the elasticity of VKM demand with respect to fuel price. Combined Eq. \ref{eq:r3} and Eq. \ref{eq:r4}, the negative $\eta_P(VKM)$ is used as a proxy measure for the rebound effect. We adopt the logarithmic model to measure the short-term direct rebound effect for road passenger transport, then we estimate the corresponding long-term direct rebound effect.

\subsection{The elasticity model}\label{section_elas}

Following most studies in the literature (e.g. Small and Van Dender, 2007; Zhang et al., 2015), we choose the logarithmic equations to investigate the size of elasticity of VKM demand with respect to fuel price. The most common assumption in the literature is that fuel price and income are the only explanatory variables for VKM demand (e.g., Alves and Bueno, 2003; Akinboade et al., 2008; Sene, 2012). However, considering that the number of vehicles explains some degree of the demand for travel, in this paper we choose price level, income level and vehicle stock as the determinants for passenger travel following by Odeck and Johansen (2016). After adding the time-lagged VKM, we take the logarithmic operation to all variables before regression in Eq.(\ref{eq:r5}), where $VKM_t$ refers to per capita demands for travel; $Y_t$ is real income per capita; $P_t$ is the real price of fuel; $V_t$ is vehicle stock; $VKM_{t-i-1}$ is the time-lagged VKM; the vector $\Lambda$ are the parameters to be estimated; and $\epsilon_t$ is residuals for VKM demand, at time $t$\footnote{Here we use per capita road passenger turnover to presents the demand for travel. The real income refers to the disposable income in the whole country. We use the 0\# diesel price to probe how the price level influences VKM. Because the vehicles of road passenger transport in China are heavy and mainly burn diesel (Zhang et al., 2015). Considering the statistical standard of the road passenger turnover in China, here $V_t$ refers to road commercial passenger vehicles. }.

\begin{equation} \label{eq:r5}
 \text{ln} VKM_t=\lambda_0+\lambda_Y \text{ln} Y_t+\lambda_P \text{ln} P_t+\lambda_V \text{ln} V_t+\sum_{i=0}^{q}\lambda_{vkmi} \text{ln} VKM_{t-i-1}+\epsilon_t
\end{equation}

We find that the result of one-ordered lagged ln$VKM$ is not significant through Eq.(\ref{eq:r5}), using the time-series data from 1986--2014. Then we use the two-lagged ln$VKM$, namely, ln$VKM_{t-2}$. Finally, the VKM demand equation can be written as follow:

\begin{equation} \label{eq:r6}
 \text{ln} VKM_t=\lambda_0+\lambda_Y \text{ln} Y_t+\lambda_P \text{ln} P_t+\lambda_V \text{ln} V_t+\lambda_{vkm} \text{ln} VKM_{t-2}+\epsilon_t
\end{equation}

where the meaning of the variables and parameters are the same as in Eq.(\ref{eq:r5}). In this way, $-\lambda_P$ and $-\frac{\lambda_P}{1-\lambda_{vkm}}$ are the size of short-term and long-term direct air pollution rebound effect for road passenger transport in China.

\subsection{Data}\label{section_data}

By collecting the time-series data from 1986 to 2014, we explore the existence of direct air pollution rebound effect for road passenger transport in China and estimate the magnitudes of short-term and long-term direct rebound effects, respectively. The utilized data are all the annual average data. National Bureau of Statistics of the People's Republic of China, which has collected official statistics on Chinese society\footnote{ \url{http://www.stats.gov.cn/}}, provides macroeconomic data, such as disposable income, population, vehicle stock and road passenger turnover. Diesel price data is obtained from ``Price Yearbook of China" (Editorial Department of Price Yearbook of China, 1989-2015) and ``Price Statistical Yearbook of China" (National Bureau of Statistics of China, 1988-1989). Variables that required conversion to their per capita forms are divided by the total annual population. Descriptive statistics of all variables are shown in Table \ref{tab2}, including the logarithmic forms of vehicles kilometers traveled per capita, the disposable income per capita, diesel price and vehicle stock.

\begin{table}[h!]
\centering
\begin{threeparttable}[b]
\captionsetup{font={scriptsize}}
\caption{Descriptive statistics of the variables in China, 1986--2014} \label{tab2}
\scriptsize
\begin{tabular}{p{2.5cm}p{2cm}p{2cm}p{2cm}p{2cm}}
\hline
Variable      &lnVKM     &lnY      &lnP      &lnV      \\ \hline	
Minimum       &2.2656    &2.7330   &3.0336	 &4.9571   \\
Maximum       &3.1348    &4.3126   &3.9354   &6.6426   \\
Mean          &2.7032    &3.5524   &3.4951   &5.6985   \\
Std. Dev.     &0.2558    &0.4790   &0.2901   &0.5849   \\
Skewness      &-0.1121   &-0.1432  &0.0843   &0.0350   \\
Kurtosis      &1.8700    &1.9057   &1.8197   &1.3909   \\
Observations  &29        &29       &29       &29       \\ \hline
\end{tabular}
\end{threeparttable}
\end{table}

\section{Empirical results and discussions}\label{section_results}

\subsection{Unit root test and cointegration test}\label{section_test}

A specific issue regarding the data's stationarity properties must be considered using time-series data. If two time-dependent variables follow a common trend that cause them to move in the same direction, it is possible to observe a significant correlation between them, even if there is no ``true" association. This potential problem with time-series data can lead to a spurious regression. To avoid the mistake resulting from spurious regression problems, all the variables are checked for their stationarity properties. We employ both the Dickey and Fuller (1979) test and the Phillips and Perron (1988) test to determine the presence of a unit root. Table \ref{tab3} presents the tests for the stationarity of the variables.

\begin{table}[h!]
\centering
\begin{threeparttable}[b]
\captionsetup{font={scriptsize}}
\caption{DF and PP test for the presence of unit root in level and differenced variables} \label{tab3}
\scriptsize
\begin{tabular}{p{2.3cm}p{2.7cm}p{2.3cm}p{2.3cm}p{2.3cm}}
\hline
Variable                &DF test   &1\% critical value   &5\% critical value    &10\% critical value\\ \hline
$lnVKM$	                &-1.576    &-3.730    &-2.992     &-2.626  \\
$lnP$                   &-0.681    &-3.730    &-2.992     &-2.626  \\
$lnY$                   &-0.994    &-3.730    &-2.992     &-2.626  \\
$lnV$                   &-1.293    &-3.730    &-2.992     &-2.626  \\
$\Delta lnVKM$          &-5.190***    &-3.736    &-2.994     &-2.628   \\
$\Delta lnP$            &-5.545***    &-3.736    &-2.994     &-2.628  \\
$\Delta lnY$            &-2.800*    &-3.736    &-2.994     &-2.628  \\
$\Delta lnV$            &-4.117***    &-3.736    &-2.994     &-2.628   \\\hline

Variable   &PP test   &1\% critical value   &5\% critical value    &10\% critical value\\ \hline
$lnVKM$	                &-1.643    &-3.730    &-2.992     &-2.626  \\
$lnP$                   &-0.667    &-3.730    &-2.992     &-2.626  \\
$lnY$                   &-0.791    &-3.730    &-2.992     &-2.626  \\
$lnV$                   &-1.376    &-3.730    &-2.992     &-2.626  \\
$\Delta lnVKM$	        &-5.193***    &-3.736    &-2.994     &-2.628	 \\
$\Delta lnP$            &-5.611***    &-3.736    &-2.994     &-2.628    \\
$\Delta lnY$            &-2.919*    &-3.736    &-2.994     &-2.628     \\
$\Delta lnV$            &-4.117***    &-3.736    &-2.994     &-2.628     \\
Notes: &All variables are I(1)\tnote{a} \\ \hline

\end{tabular}
\begin{tablenotes}
   \item[a] *** indicates the significance at 1\% level. * indicates the significance at 10\% level. $\Delta lnVKM$ denotes the first-order difference of $lnVKM$, with the similar meaning to other variables.
\end{tablenotes}
\end{threeparttable}
\end{table}

As is shown, all the variables are nonstationary at various levels for the reason that all results of DF test and PP test cannot reject the null hypothesis at the 10\% significance level. Then we test the stationarity of the first order difference of each variable, and the DF and PP tests indicate that the first differences exceed the critical value for all the variables, which indicates that all the variables are stationary in first differences, i.e., all the series are I(1). These results allow us to conduct the cointegration test to estimate whether a long-term equilibrium relationship exists among these variables, which is presented in Table \ref{tab4}. According to the Johansen (1988) test, the results of the trace statistic and the max statistic both imply that none cointegration relationship is rejected at the 5\% significance level, which means the existence of one long-run relationship. Hence, we can conclude that the long-run relationship does exist among vehicles kilometers traveled per capita, the disposable income per capita, diesel price and vehicle stock.

\begin{table}[h!]
\centering
\begin{threeparttable}[b]
\captionsetup{font={scriptsize}}
\caption{Results of Johansen test for cointegration} \label{tab4}
\scriptsize
\begin{tabular}{cccccc}
\hline
Rank             &LL        &Trace statistic   &5\% critical value     &Max statistic     &5\% critical value  \\ \hline
0	             &192.45    &64.93             &54.64                  &36.37             &30.33   \\
1                &210.64    &28.56             &34.55                  &16.23             &23.78     \\
2                &218.75    &12.33             &18.17                  &7.95              &16.87      \\
3                &224.92    &4.38              &3.74                   &4.38              &3.74       \\  \hline

\end{tabular}
\end{threeparttable}
\end{table}

\subsection{Rebound effect estimation}\label{section_OLS}

Since the variables are found to be cointegrated, we develop logarithmic regression model to estimate the coefficients according to Eq.(\ref{eq:r6}), which is reported in Table \ref{tab5}. The adjusted R-squared value is relatively high. The estimated coefficients are statistically significant. We can identify the following findings.

\begin{table}[h!]
\centering
\begin{threeparttable}[b]
\captionsetup{font={scriptsize}}
\caption{OLS estimation of the logarithmic regression model} \label{tab5}
\scriptsize
\begin{tabular}{p{4cm}p{2cm}p{2cm}p{2cm}p{2cm}}
\hline
Dependent variable      &$lnVKM_t$         &          &             &         \\   \cline{2-5}
Explanatory variables   &Coefficient   &SE         &t-statistic  &P value   \\ \hline
$lnP_t$                 &0.4105        &0.1793     &2.29         &0.032**     \\
$lnY_t$                 &0.6023        &0.2119     &2.84         &0.009***     \\
$lnV_t$                 &0.0644        &0.0248     &2.60         &0.016**      \\
$lnVKM_{t-2}$           &-0.6690       &0.3462     &-1.93        &0.066*      \\
$\lambda_0$             &0.5375        &0.3074     &1.75         &0.094*     \\
Test diagnostic                                                            \\
Adjusted R-squared            &0.96                                        \\\hline

\end{tabular}
\begin{tablenotes}
   \item *, **, *** denote the significance at 10\%, 5\% and 1\% level, respectively.
\end{tablenotes}
\end{threeparttable}
\end{table}

First, from the direct meaning of estimated coefficients, the static elasticities of travel demand (VKM) with respect to fuel price, income and vehicle stock are 0.4105, 0.6023 and 0.0644, respectively. The elasticity of fuel price and vehicle stock are significant at the 5\% significance level. The result of income is significant at the 1\% level. These results imply that an increase in the fuel price of 10\% would increase travel demand per capita by 4.105\%; an increase of disposable income per capita of 10\% would cause an increase in travel distance per capita of 6.023\%; and the travel demand per capita would increase 0.644\% if vehicle stock increases by 10\%. The magnitude of income elasticity is close to the previous studies. For instance, Zhang et al. (2015), based on data of 30 provinces from 2003 to 2012, estimate that the elasticity of passenger kilometers with respect to gross domestic product per capita (PGDP) is 0.7907 in whole China. The result of vehicle stock is relatively smaller than the income and fuel price. One reason may be that in this model we mainly use the road commercial passenger vehicles. Although the stock has an increase during the last thirty years, the growth rate of population in China is much bigger than that of vehicle stock.

Attention is drawn to a strong difference: some studies estimate the elasticity of travel demand with respect to fuel price of a negative value (e.g., Zhang et al., 2015; Odeck and Johansen, 2016; Barla et al., 2009), which means that when the fuel prices increase, the travel demand will decrease. However, our result is different from these results, which is reasonable in this paper. We mainly study the road passenger transport, and use road passenger turnover per capita as the proxy. When the diesel price increases, generally the gas price also increases as well. So the cost of taking commercial vehicles is relatively cheaper than taking private cars, resulting in the residents choosing commercial vehicles more. Furthermore, this model is based on the time-series data. China experiences a dramatic increase in the last thirty years. The growth rate of diesel price is largely smaller than the residents' travel demand. So from the result of this model, we find that when the diesel price increases, the travel demand also increases.

Second, this paper most concerns the existence of air pollution rebound effect for road passenger transport sector. From the results of Table \ref{tab5} and Eqs.(\ref{eq:r3}) and (\ref{eq:r4}), the short-term PRE can be estimated as -1.4105. This negative estimation means that the direct air pollution rebound effect does not exist in road passenger transport sector of whole China in the short-term during 1986--2014 based on the Table \ref{tab1}. The harmful gases emissions reduction is more than 1\% in the short-term when the fuel efficiency of vehicles improves by 1\% based on the unchanged emission factors. This result implies that the policies that control air pollution from transport sector are effective. The policies not only achieve the initial emissions reduction goal, but also exceed expectations.

In addition, the corresponding long-run PRE are obtained as -1.246. The negative result also implies direct PRE does not exist in the long-term during 1986--2014. However, the long-run PRE is smaller than the short-term, which means the effect of harmful emissions reduction declines than the short-term. This result puts forward a new question that whether the PRE will occur for a long time. Considering the existence of direct energy rebound effect for transport sector in developed countries (e.g., Odeck and Johansen, 2016; Hymel et al., 2010; Barla et al., 2009), it is necessary to study this problem further in future.

\section{Conclusion}\label{section_conclu}

This study, to the best of our knowledge, is the first attempt to explore whether there is direct air pollution rebound effect for road passenger transport in China, based on time-series data from the period 1986--2014. Our empirical results indicate that direct PRE does not exist in road passenger transport sector of whole China during 1986--2014. The policies, which aim to reduce harmful emissions of transport sector, are not only fully achieve the expected benefits, but also exceed the expectations. The results imply that improving fuel efficiency of vehicles is a useful policy option for decreasing the transport energy use, resulting in the reduction of harmful gases emissions.

Our empirical study show that the effect of harmful emissions reduction in the long-term declines than the short-term during 1986--2014. With China's development, the air pollution of transport sector in China may be occur rebound effect in future. The policy makers should consider the possibility of rebound effect to avoid overestimating harmful emissions reduction achieved by implementing some policies for transport sector.

Although there is no study on the air pollution rebound effect, there are studies about the energy rebound effect for transport in China. Wang et al. (2012) and Zhang et al. (2015) both find that there exists direct energy rebound effect for transport. However, their results are very different from each other. Wang et al. (2012), based on the data of 28 provinces during 1994--2009, find that the average rebound effect for passenger transport by urban households is around 96\% by employing the LA-AIDS model, indicating that the majority of the expected reduction in transport energy consumption from efficiency improvement could be offset. Zhang et al. (2015), based on the data of 30 provinces during 2003--2012, find that the average sizes of short-term and long-term rebound effect are 25.53\% and 26.56\% in the whole country through dynamic panel data model. This difference may be on account of the different methods and observations (Zhang et al., 2015). This prompts our further research can be conducted to estimate the different regions of China and use the panel data to get more observations.

Furthermore, as for the future work, further research can be conducted to combine the gasoline and diesel together, and choose more factors influencing travel demand to downsize related bias as much as possible.


\end{onehalfspace}

\clearpage
\newpage
\section*{References}

{\footnotesize

Alves, D.C.O., da S. Bueno, R.D.L., 2003. Short-run, long-run and cross elasticities of gasoline demand in Brazil. Energy Economics, 25(2): 191--199.

Akinboade, O.A., Ziramba, E., Kumo, W.L., 2008. The demand for gasoline in South Africa: An empirical analysis using co-integration techniques. Energy Economics, 30(6): 3222--3229.

Beelen, R., Hoek, G., van den Brandt, P.A., Goldbohm, R.A., Fischer, P., Schouten, L.J., Armstrong, B., Brunekreef, B., 2008. Long-term exposure to traffic-related air pollution and lung cancer risk. Epidemiology 19, 702--710.

Brunekreef, B., Beelen, R., Hoek, G., Schouten, L., Bausch-Goldbohm, S., Fischer, P., Armstrong, B., Hughes, E., Jerrett, M., v.d. Brandt, P.A., 2009. Effects of long-term exposure to traffic-related air pollution on respiratory and cardiovascular mortality in the Netherlands: the NLCS-AIR study. Research Report (Health Effects Institute) 139, 5--71.

Barla, P., Lamonde, B., Miranda-Moreno, L.F., Boucher, N., 2009. Traveled distance, stock and fuel efficiency of private vehicles in Canada: price elasticities and rebound effect. Transportation, 36(4): 389--402.

Berkhout, P.H., Muskens, J.C., Velthuijsen, J.W., 2000. Defining there bound effect. Energy Policy 28, 425--432.

Chen, S.M., He, L.Y., 2014. Welfare loss of China's air pollution: How to make personal vehicle transportation policy. China Economic Review 31, 106--118.

Dickey, D.A., Fuller, W.A., 1979. Distribution of the estimators for autoregressive time series with a unit root. Journal of the American statistical association, 74(366a): 427--431.

Editorial Department of Price Yearbook of China, 1989--2015. Price Yearbook of China. Price Yearbook of China Press, Beijing.

Frondel, M., Peters, J., Vance, C., 2008. Identifying the rebound: evidence from a German household panel. Energy Journal 29, 145--163.

Greening, L.A., Greene, D.L., Difiglio, C., 2000. Energy efficiency and consumption--the rebound effect--a survey. Energy Policy 28, 389--401.

Hoek, G., Brunekreef, B., Goldbohm, S., Fischer, P., van den Brandt, P.A., 2002. Association between mortality and indicators of traffic-related air pollution in the Netherlands: a cohort study. The Lancet 360, 1203--1209.

Hymel, K.M., Small, K.A., van Dender, K., 2010. Induced demand and rebound effects in road transport. Transportation Research Part B: Methodological, 44(10): 1220-1241.

Hymel, K.M., Small, K.A., 2015. The rebound effect for automobile travel: asymmetric response to price changes and novel features of the 2000s. Energy Economics, 49: 93--103.

Jevons, W.S., 1866. The Coal Question, 2nd ed. Macmillan and Company, London.

Johansen, S., 1988. Statistical analysis of cointegration vectors. Journal of economic dynamics and control, 12(2): 231--254.

Khazzoom, J.D., 1980. Economic implications of mandated efficiency in standards for household appliances. The Energy Journal 1(4): 21--40.

Kunzli, N., Kaiser, R., Medina, S., Studnicka, M., Chanel, O., Filliger, P., Herry, M., Horak, F., Puybonnieux-Texier, V., Quenel, P., Schneider, J., Seethaler, R., Vergnaud, J.C., Sommer, H., 2000. Public-health impact of outdoor and traffic-related air pollution: an European assessment. The Lancet 356, 795--801.

National Bureau of Statistics of China, 1988--1989. Price Statistical Yearbook of China. China Statistics Press, Beijing.

National Bureau of Statistics of China, 2006--2015. China Statistical Yearbook. China Statistics Press, Beijing.

Odeck, J., Johansen, K., 2016. Elasticities of fuel and traffic demand and the direct rebound effects: An econometric estimation in the case of Norway. Transportation Research Part A: Policy and Practice, 83: 1--13.

Phillips, P.C.B., Perron, P., 1988. Testing for a unit root in time series regression. Biometrika, 75(2): 335--346.

Samet, J.M., 2007. Traffic, air pollution, and health. Inhalation Toxicology 19, 1021--1027.

Small, K.A., Van Dender, K., 2007. Fuel efficiency and motor vehicle travel: the declining rebound effect. The Energy Journal, 28: 25--51.

Sorrell, S., Dimitropoulos, J., 2008. The rebound effect: Microeconomic definitions, limitations and extensions. Ecological Economics, 65(3): 636--649.

Sorrell, S., Dimitropoulos, J., Sommerville, M., 2009. Empirical estimates of the direct rebound effect: A review. Energy policy, 37(4): 1356--1371.

Sene, S.O., 2012. Estimating the demand for gasoline in developing countries: Senegal. Energy economics, 34(1): 189--194.

Van Rooij, B., 2006. Implementation of Chinese environmental law: regular enforcement and political campaigns. Development and Change 37(1), 57--74.

World Bank, 2007. World development indicators 2007. Washington, DC: World Bank. \url{http://siteresources.worldbank.org/DATASTATISTICS/Resources/table3_13.pdf}

Weichenthal, S., Kulka, R., Dubeau, A., Martin, C., Wang, D., Dales, R, 2011. Traffic related air pollution and acute changes in heart rate variability and respiratory function in urban cyclists. Environmental Health Perspectives 119(10), 1373.

Wang, H., Zhou, P., Zhou, D.Q., 2012. An empirical study of direct rebound effect for passenger transport in urban China. Energy Economics, 34(2): 452--460.

World Bank, 2016. World development indicators 2016. Washington, DC: World Bank.

Yang, G.H., Wang, Y., Zeng, Y.X., Gao, G.F., Liang, X.F., Zhou, M.G., Wan, X., Yu, S.C., Jiang, Y.H., Naghavi, M., Vos, T., Wang, H.D., Alan D.L., JL. Murray, C., 2013. Rapid health transition in China, 1990--2010: findings from the global burden of disease study 2010. The Lancet 381(9882), 1987--2015.

Yang, S., He, L.Y., 2016. Fuel demand, road transport pollution emissions and residents' health losses in the transitional China. Transportation Research Part D: Transport and Environment 42, 45--59.

Zhang, Y.J., Peng, H.R., Liu, Z., Tan W.P., 2015. Direct energy rebound effect for road passenger transport in China: A dynamic panel quantile regression approach. Energy Policy, 87: 303--313.



\end{document}